\documentclass[
aps,
prd,
10pt,
showpacs,
amsmath,
amssymb,
twocolumn,
nofootinbib,
nobibnotes,
preprintnumbers
]{revtex4-1}
\usepackage{float}
\usepackage{mathrsfs,amsfonts}
\usepackage{mathtools}
\usepackage{bm}
\usepackage{tensor}
\usepackage{nicefrac}
\usepackage{url}
\usepackage{xcolor}
\usepackage[hyperindex,breaklinks,pdfusetitle]{hyperref}
\let\Tr\undefined

\DeclareMathOperator{\tr}{tr}
\DeclareMathOperator{\Tr}{Tr}

\begin{document}
\allowdisplaybreaks[1]
\title{Quantum effective action for degenerate vector field theories}
\author{Michael S. Ruf}
\email{michael.ruf@physik.uni-freiburg.de}
\author{Christian F. Steinwachs}
\email{christian.steinwachs@physik.uni-freiburg.de}

\affiliation{Physikalisches Institut, Albert-Ludwigs-Universit\"at Freiburg,\\
Hermann-Herder-Stra\ss e~3, 79104 Freiburg, Germany}
%

\begin{abstract}
We calculate the divergent part of the one-loop effective action in curved spacetime for a particular class of second-order vector field operators with a degenerate principal part. The principal symbol of these operators has the structure of a longitudinal projector. In this case, standard heat-kernel techniques are not directly applicable. We present a method which reduces the problem to a non\-degenerate scalar operator for which standard heat-kernel techniques are available. Interestingly, this method leads to the identification of an effective metric structure in the longitudinal sector. The one-loop divergences are compactly expressed in terms of invariants constructed from this metric.  
\end{abstract}


\pacs{04.60.-m; 04.62.+v; 11.10.Gh; 11.15.-q; 98.80.Qc}
\maketitle


\section{Introduction}\label{SecIntro}
Perturbative calculations in quantum field theory, especially in curved spacetime, are efficiently performed by a combination of the background field method and the heat-kernel technique \cite{DeWitt1965,Atiyah1973,Gilkey1974,Gilkey1975,Abbott1981,Abbott1982,Ichinose1982,Jack1984,Barvinsky1985,Buchbinder1992,Avramidi2000,Vassilevich2003,Parker2009}. Major advantages are manifest covariance in each step of the calculation as well as universality in the sense that the formalism can be applied to fields of arbitrary spin and internal bundle structure.
For the minimal second-order operator, a closed algorithm for the calculation of the one-loop divergences, proposed by DeWitt, is available \cite{DeWitt1965}. For more general nonminimal and higher-order operators a generalization of DeWitt's algorithm was developed by Barvinksy and Vilkovisky \cite{Barvinsky1985}. The main idea of the generalized Schwinger-DeWitt technique is to reduce the calculation of more complicated operators to the known case of the minimal second-order operator by an iterative procedure which is based on the expansion of the Greens function around the principal part of the operator. However, in the case where the principal part is degenerate, direct application of the generalized Schwinger-DeWitt method fails \cite{Barvinsky1985,Ruf2018,Ruf2018b}.
In gauge theories, not only the principal part of the associated fluctuation operator is degenerate; the gauge symmetry implies that the total operator is degenerate. The gauge degeneracy can be removed by a proper gauge fixing which, in general, not only removes the degeneracy of the total operator but at the same time removes the degeneracy of the principal part. In particular, in many cases the gauge freedom is sufficient to choose a particular simple, minimal gauge. In these cases the generalized Schwinger-DeWitt technique becomes applicable again. 

There are, however, many interesting models---gauge as well as nongauge theories---which lead to fluctuation operators for which the degeneracy of the principal part cannot be removed. This happens, for example, in softly broken gauge theories where no gauge fixing is available and in higher-derivative theories, where only some of the degrees of freedom are propagating with higher derivatives. Particularly relevant models where this is the case are $f(R)$ gravity and the generalized Proca field in curved spacetime, which both provide the basis of many important cosmological applications. The renormalization structure of these models and their one-loop divergences on an arbitrary background were investigated in Refs.~\cite{Ruf2018, Ruf2018b}.

For the explicit calculation of the one-loop divergences, various methods were developed to overcome the difficulties  associated with the degenerate principal part.
The degeneracy of the principal part is inextricably linked with its nonminimal derivative structure. The simplest class of operators with nonminimal principal part are vector field operators. A systematic classification of vector field operators according to their degeneracy structure has been developed in Ref.~\cite{Ruf2018b}. The present article completes this classification by adding a class of vector field operators with nonminimal principal part without a Laplacian. Vector field theories which lead to such a fluctuation operator feature nonwavelike equations of motion; see Ref.~\cite{Avramidi2001} for a discussion of these operators in the context of the heat kernel. The heat-kernel technique for more ``exotic'' operators has also been studied previously in Ref.~\cite{Fulling1992}.

In this article, we derive the divergent part of the one-loop effective action for the longitudinal vector field operator in a closed form. 
A particularly interesting feature is that the result is compactly represented in terms of geometrical invariants constructed from an additional metric structure which emerges naturally from the potential part of the vector field operator.

The paper is structured as follows: In Sec.~\ref{Sec:LongVector}, we introduce the class of degenerate vector field operators considered in this article. In Sec.~\ref{Sec:One-loop}, the one-loop divergences are calculated in a closed form. In Sec.~\ref{Sec:CrossCheck} we check our result by an alternative method of calculation for a special case. We conclude in Sec.~\ref{Sec:Conclusion} and give a brief outline of future generalizations of the obtained results.    
\vfill
 
\section{The longitudinal vector field operator}\label{Sec:LongVector}
The class of degenerate vector field operators considered in this article arise, for example, from the following Euclidean action functional for a vector field $A^{\mu}$,
\begin{align}
S[A]=\frac{1}{2}\int_{\mathcal{M}}\mathrm{d}^4x g^{1/2}\left[\left(\nabla_{\mu}A^{\mu}\right)^2+M_{\mu\nu}A^{\mu}A^{\nu}\right].\label{action}
\end{align}
Here, $\nabla_{\mu}$ is the torsion-free covariant derivative compatible with the spacetime metric $g_{\mu\nu}$ on the $d=4$ dimensional manifold $\mathcal{M}$. If not otherwise indicated, derivatives $\nabla_{\mu}$ always act on everything to their right. The generalized mass tensor $M_{\mu\nu}$ is assumed to be symmetric and positive definite.
The Hessian $H_{\mu\nu}(\nabla)$ of Eq.~\eqref{action} defines a second-order differential operator,
\begin{align}
H_{\mu\nu}(\nabla^{x})\delta(x,y)={}&\frac{\delta^2 S[A]}{\delta A^{\mu}(x)\delta A^{\nu}(y)}\nonumber\\
={}&g^{1/2}\left(-\nabla_{\mu}^{x}\nabla_{\nu}^{x}+M_{\mu\nu}\right)\delta(x,y)\,.\label{Hessvec}
\end{align} 
The superscripts, which we will suppress in what follows, indicate that the covariant derivatives act at the point $x$. The delta function is assumed to be a scalar density of zero weight in its first argument and of unit weight at its second argument. The natural metric on the space of vector fields is given by
\begin{align}
\gamma_{\mu\nu}\coloneqq g^{1/2}g_{\mu\nu}\,.\label{confmetricvec}
\end{align}  
Note that $\gamma_{\mu\nu}$ includes the density factor $g^{1/2}$. It defines the inner product on the space of vector fields
\begin{align}
\langle A,A\rangle_{1}\coloneqq\int_{\mathcal{M}}\mathrm{d}^4x\,\gamma_{\mu\nu}A^{\mu}A^{\nu}\,.\label{InnerProduct}
\end{align}
We define the fluctuation operator with natural index positions as
\begin{align}
\tensor{F}{^{\mu}_{\nu}}(\nabla)\coloneqq \left(\gamma^{-1}\right)^{\mu\rho}H_{\rho\nu}(\nabla)=-\nabla^{\mu}\nabla_{\nu}+\tensor{M}{^{\mu}_{\nu}}\,,\label{FOPComp}
\end{align}
where $\left(\gamma^{-1}\right)^{\mu\nu}=g^{-1/2}g^{\mu\nu}$ is the reciprocal of $\gamma_{\mu\nu}$ and $\tensor{M}{^{\mu}_{\nu}}=g^{\mu\rho}M_{\rho\nu}$.
The linear operator $\mathbf{F}$ naturally acts as the matrix $\tensor{F}{^\mu_\nu}$ on the space of vectors $A^{\mu}$. We denote such linear operators in boldface and only resort to the explicit components $\tensor{F}{^\mu_\nu}$ if necessary.
In this compact notation, the vector operator \eqref{FOPComp} reads
\begin{align}
\mathbf{F}(\nabla)=\bm{\nabla}^{\dagger}\bm{\nabla}+\mathbf{M},\label{FOP}
\end{align}
with $\bm{\nabla}^{\dagger}$ being the adjoint of $\bm{\nabla}$ with respect to the inner product \eqref{InnerProduct}. The operator \eqref{FOPComp} has a degenerate principal part
\begin{align}
\mathbf{D}(\nabla)\coloneqq \bm{\nabla}^{\dagger}\bm{\nabla}\,.\label{PPart}
\end{align}
The associated principal symbol, obtained by the formal replacement $\bm{\nabla}\to\mathrm{i} \mathbf{n}$ with a constant vector $\mathbf{n}$, has the structure of the projector $\mathbf{\Pi}_{\parallel}=\bm{n}^{\dagger}\bm{n}/n^2$ onto the longitudinal mode of $A^{\mu}$ and $n^2\coloneqq n_{\mu}n^{\mu}$:
\begin{align}
\mathbf{D}(n)=n^2\mathbf{\Pi}_{\parallel},\qquad \det\mathbf{D}(n)=0\,.\label{Projector}
\end{align}
In other words, the operator \eqref{FOP} has zero eigenvalue eigenvectors, and the associated Greens function cannot be obtained as a perturbative expansion in $\mathbf{M}$, 
\begin{align}
\frac{\mathbf{1}}{D+M}=\frac{\mathbf{1}}{D}-\frac{\mathbf{1}}{D}\mathbf{M}\frac{\mathbf{1}}{D}+\cdots.
\end{align}
The particular type of degenerate vector operator \eqref{FOP} where the principal symbol has the structure of a longitudinal projector \eqref{Projector} has not been included in the classification scheme of Ref.~\cite{Ruf2018b} and is discussed in what follows.

\section{One-loop divergences}\label{Sec:One-loop}
The divergent part of the one-loop effective action for the operator \eqref{FOP} is obtained from
\begin{align}
\Gamma_{1}^{\mathrm{div}}=\frac{1}{2}\left.\Tr_{1}\ln\mathbf{F}\right|^{\mathrm{div}}\,.\label{DefGam1div}
\end{align}
Here, the functional trace $\Tr_{1}$ is performed over vector fields $A^{\mu}$.
By a sequence of formal manipulations, we reduce the vector trace to a scalar trace, corresponding to the propagating longitudinal mode. The divergent part of the one-loop effective action then reads
\begin{align}
\Gamma_{1}^{\mathrm{div}}
={}&\frac{1}{2}\left.\Tr_{1}\ln\left(\bm{\nabla}^{\dagger}\bm{\nabla}+\mathbf{M}\right)\right|^{\mathrm{div}}\nonumber\\
={}&\frac{1}{2}\left.\Tr_{1}\ln\mathbf{M}\right|^{\mathrm{div}}+\frac{1}{2}\left.\Tr_{1}\ln\left(\mathbf{1}+\bm{\nabla}^{\dagger}\bm{\nabla}\mathbf{M}^{-1}\right)\right|^{\mathrm{div}}\nonumber\\
={}&\frac{1}{2}\sum_{n=1}^{\infty}\frac{(-1)^{n-1}}{n}\left.\Tr_1\left(\bm{\nabla}^{\dagger}\bm{\nabla}\mathbf{M}^{-1}\right)^{n}\right|^{\mathrm{div}}\nonumber\\
={}&\frac{1}{2}\left.\Tr_{0}\ln\left(1+\bm{\nabla}\mathbf{M}^{-1}\bm{\nabla}^{\dagger}\right)\right|^{\mathrm{div}}\nonumber\\
={}&\frac{1}{2}\left.\Tr_{0}\ln \mathbf{F}_{\mathrm{s}}\right|^{\mathrm{div}}\label{Gamm1div}\,.
\end{align}
Here we have used the cyclicity of the trace and the rule 
\begin{align}
\left.\Tr\log\left(\mathbf{L}_1\mathbf{L}_2\right)\right|^{\mathrm{div}}=\left.\Tr\log\mathbf{L}_1\right|^{\mathrm{div}}+\left.\Tr\log\mathbf{L}_2\right|^{\mathrm{div}}
\end{align}
for two linear operators $\mathbf{L}_1$ and $\mathbf{L}_2$. We have also used the fact that the functional trace over the generalized mass tensor $\mathbf{M}$ does not contribute to the divergent part ${\left.\Tr_{1}\ln\mathbf{M}\right|^{\mathrm{div}}=0}$. In the last equality, we have defined the formally self-adjoint scalar operator 
\begin{align}
F_{\mathrm{s}}(\nabla)\coloneqq
 -\nabla_{\mu}\left(\tilde{g}^{-1}\right)^{\mu\nu}\nabla_{\nu}+m^2\,,\label{ScalarOp}
\end{align}
where we have introduced a new metric 
\begin{align}
\tilde{g}_{\mu\nu}\coloneqq\frac{M_{\mu\nu}}{m^2}\,.\label{gtilde}
\end{align}
Here, $m$ is an auxiliary constant parameter with the dimension of mass, introduced to make $\tilde{g}_{\mu\nu}$ dimensionless. Thus, the formal manipulations in Eq.~ \eqref{Gamm1div} by which the vector trace is converted into a scalar trace, naturally induces a second metric structure for the longitudinal scalar which is constructed by the generalized mass tensor $M_{\mu\nu}$.  
The positive definiteness of $M_{\mu\nu}$ implies that the reciprocal $\left(\tilde{g}^{-1}\right)^{\mu\nu}$ exists.
From now on, we raise and lower indices exclusively with $\left(\tilde{g}^{-1}\right)^{\mu\nu}$ and $\tilde{g}_{\mu\nu}$. 
The metric $\tilde{g}_{\mu\nu}$ uniquely defines a torsion-free, metric compatible covariant derivative $\tilde{\nabla}_{\mu}$ with the  connection
\begin{align}
\tilde{\Gamma}^{\rho}_{\mu\nu}=\frac{1}{2}\left(\tilde{g}^{-1}\right)^{\rho\sigma}\left(\partial_{\mu}\tilde{g}_{\sigma\nu}+\partial_{\nu}\tilde{g}_{\mu\sigma}-\partial_{\sigma}\tilde{g}_{\mu\nu}\right)\,.
\end{align}
The two covariant derivatives $\tilde{\nabla}_{\mu}$ and $\nabla_{\mu}$ differ by the difference tensor
\begin{align}
\delta\Gamma^{\rho}_{\mu\nu}\coloneqq{}&\tilde{\Gamma}^{\rho}_{\mu\nu}-\Gamma^{\rho}_{\mu\nu}\nonumber\\
={}&\frac{1}{2}\left(\tilde{g}^{-1}\right)^{\rho\sigma}\left(\nabla_{\mu}\tilde{g}_{\sigma\nu}+\nabla_{\nu}\tilde{g}_{\mu\sigma}-\nabla_{\sigma}\tilde{g}_{\mu\nu}\right).\label{DifferenceTensor}
\end{align}
The new metric $\tilde{g}_{\mu\nu}$ and the covariant derivative $\tilde{\nabla}_{\mu}$ suggest to define the Laplacian
\begin{align}
\tilde{\Delta}\coloneqq -\left(\tilde{g}^{-1}\right)^{\mu\nu}\tilde{\nabla}_{\mu}\tilde{\nabla}_{\nu}\,.
\end{align}
For the Laplacian acting on a scalar, we have
\begin{align}
\tilde{\Delta}={}&-\left(\tilde{g}^{-1}\right)^{\mu\nu}\tilde{\nabla}_{\mu}\tilde{\nabla}_{\nu}\nonumber\\
={}&-\left(\tilde{g}^{-1}\right)^{\mu\nu}\nabla_{\mu}\nabla_{\nu}+\left(\tilde{g}^{-1}\right)^{\mu\nu}\delta\Gamma^{\rho}_{\mu\nu}\tilde{\nabla}_{\rho}.
\end{align}
The coefficient of the last term can be written as
\begin{align}
\left(\tilde{g}^{-1}\right)^{\mu\nu}\delta\Gamma^{\rho}_{\mu\nu}=-\left[\nabla_{\mu}\left(\tilde{g}^{-1}\right)^{\mu\rho}\right]-2W^{\rho}\,,
\end{align}
where we have defined
\begin{align}
W_{\mu}\coloneqq\tilde{g}^{-1/2}\nabla_{\mu}\tilde{g}^{1/2}=\frac{1}{2}\nabla_{\mu}\ln\tilde{g},
\end{align}
with $\tilde{g}\coloneqq \det \tilde{g}_{\mu\nu}$ and $W^{\rho}=\left(\tilde{g}^{-1}\right)^{\rho\mu}W_{\mu}$.
The scalar operator \eqref{ScalarOp} then formally acquires the structure of a minimal second-order operator:
\begin{align}
F_{\mathrm{s}}= \tilde{\Delta}+2W^{\rho}\tilde{\nabla}_{\rho}+m^2\,.
\end{align}
By changing covariant derivatives $\tilde{\nabla}_{\mu}\to\mathcal{D}_{\mu}=\tilde{\nabla}_{\mu}+W_{\mu}$, we can absorb the term linear in derivatives and obtain the Laplace-type scalar operator
\begin{align}
F_{\mathrm{s}}(\mathcal{D})\coloneqq -\mathcal{D}^2+P\,,\label{ScalOp2}
\end{align}
with the scalar potential
\begin{align}
P={}& m^2 + \frac{1}{4}\left(\tilde{g}^{-1}\right)^{\mu\nu}\left(W_{\mu}W_{\nu} +2  {\nabla}_{\mu}W_{\nu}\right)\nonumber\\
={}& m^2 + \tilde{g}^{-1/4}\left(\tilde{g}^{-1}\right)^{\mu\nu} \nabla_\mu\nabla_\nu \tilde{g}^{1/4}\,.\label{ScalPot}
\end{align}
The one-loop divergences for a general minimal second-order operator of the form $-\mathcal{D}^2\mathbf{1}+\mathbf{P}$ in $d=4$ are known in closed form
\begin{align}
	\Gamma_{1}^{\mathrm{div}}={}&\frac{1}{2}\Tr\ln \left(-\mathcal{D}^2\mathbf{1}+\mathbf{P}\right)\Big|^{\mathrm{div}}\nonumber\\
	={}&-\frac{1}{32\pi^2\varepsilon}\int_{\mathcal{M}}\mathop{}\!\mathrm{d}^4x\, g^{1/2}\,\tr\,\mathbf{a}_{2}(x,x)\,,\label{1LActionMinimal}
\end{align}
where the coincidence limit of the second Schwinger-DeWitt coefficient up to total divergences is given by
\begin{align}
		\mathbf{a}_{2}(x,x)={}&\frac{1}{180}\left(\tensor{R}{_{\alpha\beta\gamma\delta}}\tensor{R}{^{\alpha\beta\gamma\delta}}-\tensor{R}{_\alpha_\beta}\tensor{R}{^\alpha^\beta}\right)\mathbf{1}\nonumber\\
	&+\frac{1}{2}\left(\mathbf{P}-\frac{1}{6}R\mathbf{1}\right)^2+\frac{1}{12}\mathbf{R}_{\alpha\beta}\mathbf{R}^{\alpha\beta}\,.\label{SDWa2Coeff}
\end{align}
The bundle curvature $\mathbf{R}_{\mu\nu}$ vanishes for a scalar field $\varphi$,
\begin{align}
\left[\mathcal{D}_{\mu},\mathcal{D}_{\nu}\right]\varphi={}&
\mathbf{R}_{\mu\nu}(\mathcal{D})\varphi=
0\,.\label{BundleCurv}
\end{align}
The one-loop divergences \eqref{DefGam1div} for the longitudinal vector field operator \eqref{FOPComp} reduce to the evaluation of the functional trace of the scalar operator \eqref{ScalOp2}. 
We obtain the final result by substituting the scalar potential \eqref{ScalPot}, the bundle curvature \eqref{BundleCurv}, and the metric \eqref{gtilde} into the general formulas \eqref{1LActionMinimal} and \eqref{SDWa2Coeff}, and by performing the internal scalar trace $\tr\mathbf{1}=1$:
\begin{align}
\Gamma_{1}^{\mathrm{div}}={}&-\frac{\chi(\mathcal{M})}{180\varepsilon}-\frac{1}{32\pi^2\varepsilon}\int_{\mathcal{M}}\mathrm{d}^4x\, \tilde{g}^{1/2}\left[\frac{1}{60}\tensor{\tilde{R}}{_\mu_\nu}\tensor{\tilde{R}}{^\mu^\nu}\right.\nonumber\\
&\left.+\frac{1}{120}\tilde{R}^2-\frac{1}{6}\tilde{R}P+\frac{1}{2}P^2\right]\,.\label{FinRes}
\end{align}
We have expressed the final result in terms of the geometrical invariants constructed from the metric $\tilde{g}_{\mu\nu}$, which is directly related to the generalized mass tensor $M_{\mu\nu}$. In Eq.~\eqref{FinRes}, we also traded the square of the Riemann tensor for the Gauss-Bonnet term
\begin{align}
\mathcal{\tilde{G}}\coloneqq{}&\tensor{\tilde{R}}{_\mu_\nu_\rho_\sigma}\tensor{\tilde{R}}{^\mu^\nu^\rho^\sigma}-4\,\tensor{\tilde{R}}{_\mu_\nu}\tensor{\tilde{R}}{^\mu^\nu}+\tilde{R}^2\,.\label{GaussBonnet}
\end{align}
The integral over the the Gauss-Bonnet density $\tilde{g}^{1/2}\mathcal{\tilde{G}}$ is equal to the Euler characteristic $\chi(\mathcal{M})$ in $d=4$,
\begin{align}
\chi(\mathcal{M})\coloneqq{}&\frac{1}{32\pi^2}\int_{\mathcal{M}}\mathrm{d}^4x\, \tilde{g}^{1/2}\mathcal{\tilde{G}}\,.
\end{align}
Since $\chi(\mathcal{M})$ is a topological invariant of the manifold $\mathcal{M}$, it is independent of the metric $\tilde{g}_{\mu\nu}$ and therefore of the generalized mass tensor $M_{\mu\nu}$.

The result in Eq.~\eqref{FinRes} shows that one-loop divergences for the longitudinal vector field operator \eqref{FOPComp} can be expressed in a closed and compact form in terms of geometrical invariants constructed from the generalized mass tensor $M_{\mu\nu}$. This might be a surprising result at first glance, as $M_{\mu\nu}$ enters the final result in a nonpolynomial way. However, the presence of the generalized mass tensor is the characteristic feature of the theory in Eq.~ \eqref{action} and the reason for the contributions of the transversal vector degrees of freedom to the one-loop divergences. Despite the longitudinal projector structure of the principal part, the generalized mass tensor induces a mixing between transversal and longitudinal degrees of freedom and distinguishes the vector theory \eqref{action} from a pure scalar field theory. This point is discussed in more detail in the next section.

\section{Cross check}\label{Sec:CrossCheck}
We perform a simple cross check of the result \eqref{FinRes} for the special case where the generalized mass tensor reduces to the ordinary mass term $M_{\mu\nu}=m^2g_{\mu\nu}$. In this case, the scalar potential \eqref{ScalPot} reduces to $P=m^2$, and the geometric invariants are defined with respect to $g_{\mu\nu}$. Consequently, the one-loop divergences \eqref{FinRes} reduce to
\begin{align}
\Gamma_{1}^{\mathrm{div}}={}&-\frac{\chi(\mathcal{M})}{180\varepsilon}-\frac{1}{32\pi^2\varepsilon}\int_{\mathcal{M}}\mathrm{d}^4x\,g^{1/2}\left[\frac{1}{60}\tensor{R}{_\mu_\nu}\tensor{R}{^\mu^\nu}\right.\nonumber\\
&\left.+\frac{1}{120}R^2-\frac{1}{6}Rm^2+\frac{1}{2}m^4\right]\,.\label{RedGenRes}
\end{align} 
This result can be obtained also in a different way. Performing the decomposition of the vector field 
\begin{align}
A^{\mu}=A^{\mu}_{\perp}+A^{\mu}_{\parallel},
\end{align}
into a transverse part $\nabla_{\mu}A^{\mu}_{\perp}=0$ and a longitudinal part $A^{\mu}_{\parallel}=g^{\mu\nu}\partial_{\mu}\varphi$ with the longitudinal scalar field $\varphi$, the action \eqref{action} reads
\begin{align}
S[A_{\perp},\varphi]={}&\int_{\mathcal{M}}\mathrm{d}^4xg^{1/2}\left[\varphi\left(\Delta^2+m^2\Delta\right)\varphi\right.\nonumber\\
&\left.+m^2g_{\mu\nu}A^{\mu}_{\perp}A^{\mu}_{\perp}\right]\,.
\end{align}
In terms of the generalized field 
\begin{align}
\phi^{A}=\left(\begin{matrix}
A^{\mu}_{\perp}\\\varphi
\end{matrix}\right),
\end{align}
the fluctuation operator acquires a block matrix form
\begin{align}
\mathbf{F}=\left(\begin{matrix}m^2\delta_{\mu}^{\nu}&0\\
0&\Delta^2+m^2\Delta\end{matrix}\right)\,.\label{OpBlock}
\end{align}
Only some of the relativistic degrees of freedom of the vector field $A^{\mu}$ are propagating with higher derivatives. Here, the longitudinal scalar $\varphi$ propagates with fourth-order derivatives, while the transversal part $A^{\mu}_{\perp}$ does not propagate. This is a result of the special projector structure of the principal symbol \eqref{Projector} and the fact that the generalized mass tensor $M_{\mu\nu}=m^2g_{\mu\nu}$ is ultralocal, covariantly constant, and has trivial index structure.
The Jacobian for the transition to the differentially constraint fields $A^{\mu}\to(A^{\mu}_{\perp},\varphi)$ is obtained from
\begin{align}
\langle A,A\rangle_{1}=\langle A_{\perp},A_{\perp}\rangle_1+\langle \varphi,\Delta\varphi\rangle_0.
\end{align}
It has the block matrix structure
\begin{align}
\mathbf{J}=\left(\begin{matrix}\delta^{\mu}_{\nu}&0\\
0&\Delta\end{matrix}\right)\,.\label{Jac}
\end{align}
The fact that the block matrices in Eqs.~\eqref{OpBlock} and \eqref{Jac} are diagonal is also a consequence of the special case ${M_{\mu\nu}=m^2g_{\mu\nu}}$. Hence no mixing between $A_{\perp}^{\mu}$ and $\varphi$ occurs. 
The divergent part of the one-loop divergences is given by
\begin{align}
\Gamma_{1}^{\mathrm{div}}=\frac{1}{2}\left.\Tr\ln\mathbf{F}\right|^{\mathrm{div}}-\frac{1}{2}\left.\Tr\ln\mathbf{J}\right|^{\mathrm{div}}\,.
\end{align}
The functional traces of the block operators splits into a sum of transverse and scalar traces for the block operators corresponding to the diagonal components.
Since the transversal part is not propagating, the transversal traces do not contribute to the one-loop divergences. Moreover, the scalar part of the operator $\mathbf{F}$ factorizes to $\Delta(\Delta+m^2)$ and partially cancels the contribution of the scalar operator $\Delta$ from the Jacobian. The final result for the one-loop divergences of the action \eqref{action} is therefore given by that of a massive scalar field:
\begin{align}
\Gamma_{1}^{\mathrm{div}}=\frac{1}{2}\left.\Tr_0\ln\left(\Delta+m^2\right)\right|^{\mathrm{div}}\,.\label{TraceScalar}
\end{align}
Inserting the scalar operator $\Delta+m^2$ into the general formulas \eqref{1LActionMinimal} and \eqref{SDWa2Coeff}, and using the fact that ${\mathbf{R}_{\mu\nu}(\nabla)=0}$ for a scalar field $\varphi$, the result obtained from Eq.~ \eqref{TraceScalar} coincides with the reduction \eqref{RedGenRes} of the general result \eqref{FinRes} for the simple case $M_{\mu\nu}=m^2g_{\mu\nu}$. This shows again that in the general case, where the block operators \eqref{OpBlock} and \eqref{Jac} are not diagonal, the transversal vector field degrees of freedom contribute to the one-loop divergences due to the mixing with the longitudinal degrees of freedom induced by the generalized mass tensor $M_{\mu\nu}$.

\section{Conclusion}\label{Sec:Conclusion}
We calculated the one-loop divergences for a class of second-order vector field operators with degenerate principal part for which standard heat-kernel techniques are not directly applicable. By a formal manipulation of the vector trace \eqref{Gamm1div}, the calculation could be reduced to the evaluation of the functional trace for a minimal second-order scalar operator \eqref{ScalOp2}. During this procedure, an additional effective metric structure  \eqref{gtilde}, essentially given by the generalized mass tensor $M_{\mu\nu}$, arises in a natural way. The resulting one-loop divergences are expressed compactly in terms of curvature invariants constructed from this additional metric \eqref{FinRes}. Therefore, the generalized mass tensor enters the one-loop divergences in a nonpolynomial form. The origin of this rather surprising result is traced back to the particular degeneracy structure of the fluctuation operator. For a relativistic field, this happens if the principal part is degenerate but the total operator is not.
The degeneracy of the principal part, in turn, necessarily requires a particular nonminimal derivative structure. 
The situation becomes more transparent if formulated in terms of the irreducible decomposition of the field. In this case, the fluctuation operator generally becomes minimal but matrix valued, and the degeneracy manifests itself in a singular principal matrix. This means that only some components of the relativistic field propagate with higher derivatives. The standard methods are still applicable in case the different components decouple, but they fail if these components are coupled in the lower-derivative parts of the fluctuation operator.

In the case of the vector field operator considered in this article, this 
becomes evident in the context of the special case, discussed in Sec.~\ref{Sec:CrossCheck}.  For a general tensor $M_{\mu\nu}$, the matrix-valued fluctuation operator for the transversal and longitudinal components of the vector field \eqref{OpBlock} would contain off-diagonal terms in the lower-derivative part.
These off-diagonal elements lead to the aforementioned mixing between the transversal and longitudinal components. Only in case the generalized mass tensor reduces to the ordinary mass term $M_{\mu\nu}=m^2g_{\mu\nu}$, the total fluctuation operator is block diagonal and the two components decouple. A particularity of this case is that the transversal component is not propagating. A similar but complementary situation arises in the context of the generalized Proca field, where the roles are reversed and the longitudinal mode is not propagating \cite{Ruf2018b}.
In this sense, the present analysis completes the classification of vector field operators introduced in Ref.~\cite{Ruf2018b}. 

The method of Eq.~\eqref{Gamm1div} is not only restricted to second-order vector field operators but can be generalized to higher-order operators and higher-spin fields. In fact, a similar technique has been applied to the fourth-order tensor field operator arising in $f(R)$ gravity \cite{Ruf2018}. The fluctuation operator has a degenerate principal part with a product structure similar to that of the vector field operator \eqref{PPart}. 
Beside these similarities, a crucial difference to the vector field operator \eqref{FOP} is the structure of the lower-derivative part. In contrast to the generalized mass tensor, the lower-derivative part of the $f(R)$ operator is a minimal second-order operator \cite{Ruf2018}. In this case, the analogue of the procedure \eqref{Gamm1div} requires additional care \cite{Ruf2018}.

The particular degeneracy structure of the fluctuation operator discussed in this article is generically present in softly broken gauge theories. The quantization of such theories has been discussed recently in the context of vector field and tensor field models \cite{Buchbinder2007, Aashish2018,Ruf2018b}. The same type of degeneracy also arises in models of massive gravity \cite{Buchbinder2012, Hinterbichler2012, Rham2014}, cosmological Galileon models \cite{Nicolis2009, Deffayet2009, PaulaNetto2012, Saltas2017}, and generalized Proca models \cite{Heisenberg2014,Heisenberg2018,Ruf2018b}.
The identification of effective metric structures is also an essential technical feature of the heat-kernel method for anisotropic operators developed in Refs.~\cite{Nesterov2011, DOdorico2015, Barvinsky2017a}. Such operators emerge in theories without fundamental Lorentz invariance and are of particular importance for the renormalization of Ho\v{r}ava gravity \cite{Horava2009,Horava2009a,Giribet2010, LopezNacir2012,DOdorico2014, Barvinsky2016,Barvinsky2017a,Barvinsky2018}.


\acknowledgements
The work of M. S. R. is supported by the Alexander von Humboldt Foundation, in the framework of the Sofja Kovalevskaja Award 2014, endowed by the German Federal Ministry of Education and Research.

\bibliography{HKVecLongV2}{}

\end{document}